\documentclass[preprint,12pt]{elsarticle}

\usepackage{amssymb}

\journal{Journal of Magnetism and Magnetic Materials}

\begin{document}

\begin{frontmatter}



\title{High concentration ferronematics in low magnetic fields}


\author{T.~T\'oth-Katona\corref{cor1}}
\ead{tothkatona.tibor@wigner.mta.hu}

\author{P.~Salamon\corref{}}
\ead{salamon.peter@wigner.mta.hu}

\author{N.~\'Eber}
\ead{eber.nandor@wigner.mta.hu}

\address{Institute for Solid State Physics and Optics, Wigner Research Centre for Physics, Hungarian Academy of Sciences, H-1525 Budapest, P.O.Box 49, Hungary}

\author{N.~Toma\v{s}ovi\v{c}ov\'a}
\ead{nhudak@saske.sk}

\author{Z.~Mitr\'oov\'a}
\ead{mitro@saske.sk}

\author{P.~Kop\v{c}ansk\'y}
\ead{kopcan@saske.sk}

\address{Institute of Experimental Physics, Slovak Academy of Sciences, Watsonov\'a 47, 04001 Ko\v{s}ice, Slovakia}

\cortext[cor1]{Corresponding author}



\begin{abstract}
We investigated experimentally the magneto-optical and dielectric properties of magnetic-nanoparticle-doped nematic liquid crystals (ferronematics). Our studies focus on the effect of the very small orienting bias magnetic field $B_{bias}$, and that of the nematic director pretilt at the boundary surfaces in our systems sensitive to low magnetic fields. Based on the results we assert that $B_{bias}$ is not necessarily required for a detectable response to low magnetic fields, and that the initial pretilt, as well as the aggregation of the nanoparticles  play an important (though not yet explored enough) role.

\end{abstract}


\end{frontmatter}


\section{Introduction}
\label{Introduction}

The control of the orientational order of liquid crystals (LCs) by magnetic field is much less wide-spread in practise than the control by electric field. The reason for this is the relatively small anisotropy of the diamagnetic susceptibility of liquid crystals. In order to overcome this difficulty, doping of LCs with magnetic nanoparticles has been proposed theoretically long time ago   \cite{Brochard1970}. After the first experimental realization \cite{Chen1983}, the idea has been extensively tested in ferronematic suspensions of various compositions -- see e.g., \cite{Kopcansky2008,Kopcansky2010,Mertelj2013}, review articles \cite{Tomasovicova2012,Lagerwall2012}, and references therein. During these experiments an important difficulty has arisen: the aggregation of the nanoparticles \cite{Buluy2011}.

A measurable optical response to low (potentially important for applications) magnetic field has been reported only lately. A linear response has been detected in planarly oriented ferronematic samples far below the threshold of the magnetic Fr\'eedericksz transition $B_{F}$, howerver, in the presence of a weak orienting bias magnetic field ($B_{bias}\approx 2$ mT) \cite{Podoliak2011}. More recently, it has been shown that a similar response can be obtained even in the absence of $B_{bias}$ \cite{Tomasovicova2013}.

The motivation of this paper was to explore the role of $B_{bias}$, of the initial pretilt, and that of the aggregation of nanoparticles on the response of ferronematics to low magnetic fields (below $B_{F}$).

\section{Experimental}
\label{Experimental}

The thermotropic nematic 4-(trans-4'-n-hexylcyclohexyl)-isothiocyanatobenzene (6CHBT) was used as the LC matrix, which was doped either with spherical Fe$_3$O$_4$ nanoparticles having a mean diameter of about $12$~nm \cite{Tomasovicova2013}, or with single-wall carbon nanotubes functionalized with Fe$_3$O$_4$ nanoparticles (SWCNT/Fe$_3$O$_4$) \cite{Mitroova2011} in a relatively high volume concentration of $2 \times 10^{-3}$.

\begin{figure}
\centerline{\includegraphics[width=0.4\textwidth]{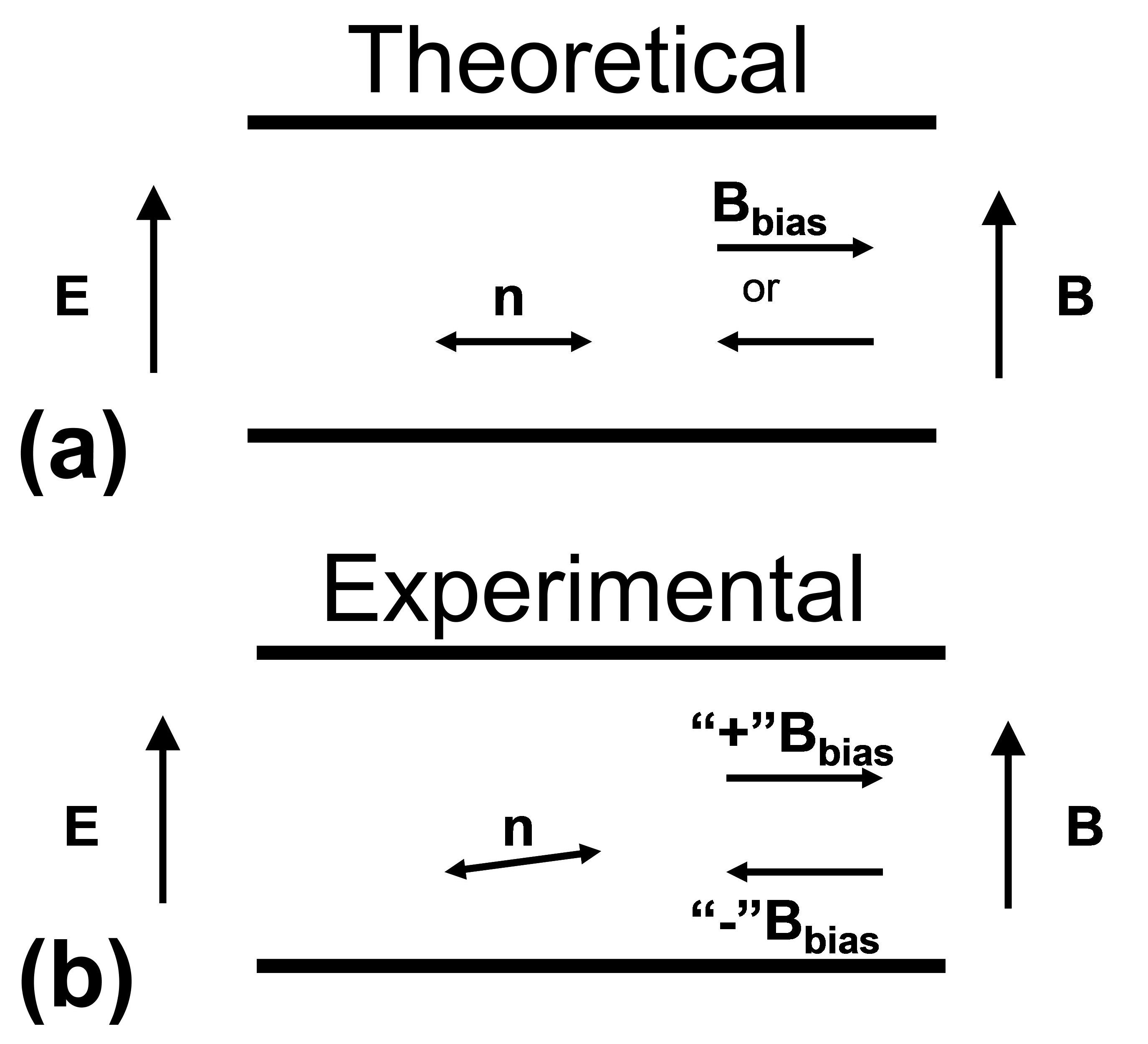}}
\caption{Schematic representation of the experimental setup: (a) the pretilt angle is neglected (theoretical); (b) the pretilt angle is nonzero (experimental). Notations: {\bf n} -- the nematic director, ${\bf B}$ -- the direction of the magnetic field,
${\bf E}$ -- the direction of the electric field, $"+"$ and $"-"$ ${\bf B}_{bias}$ -- direction(s) of the orienting bias magnetic field.}
\label{fig:Toth_Fig1}
\end{figure}

The ferronematics have been filled into $d \approx 50 \mu$m thick, planarly oriented cells. The planar  orientation was ensured by the anti-parallel rubbing of the polyimide layers coated on the inner surfaces of the two glass plates constituting the cell. The experimental setup was similar to that described in Refs. \cite{Podoliak2011,Podoliak2012}. The cells were placed in a costum-made hot-stage having a thermal stability better than $0.05^{\circ}$C. The cells could be exposed simultaneously to a magnetic induction $B$ (up to $1 \mathrm{T}$), to an electric electric field $E$, and to an orienting bias magnetic field of $B_{bias}=2\mathrm{mT}$ in an experimental geometry shown schematically in Fig.\ref{fig:Toth_Fig1}. The capacitance $C$ and the conductance $G$ were monitored by a Hioki 3522 impedance analyzer. Additionally, the setup allowed for optical studies in which the intensity of the transmitted light $I$ was measured with crossed polarizers at an orientation of $\pm 45^{\circ}$ with respect to the initial director ${\bf n}$. A laser diode emitting at $\lambda=657.3$~nm was used as a light source. The measurement control as well as the data collection was ensured by a LabVIEW program.

In the theoretical description of the planar orientation it is usually assumed that the nematic director ${\bf n}$ (the unit vector describing the orientational order of the LC) is parallel with the bounding glass plates [Fig.\ref{fig:Toth_Fig1}(a)]. In real cells, however, ${\bf n}$ encloses a small pretilt-angle with the glass plates, as shown in Fig.\ref{fig:Toth_Fig1}(b). For cells with antiparallel rubbed polyimide layers, the pretilt-angle $\theta_0$ is typically between $1^{\circ}$ and $3^{\circ}$ \cite{Qi2008}. A nonzero $\theta_0$ breaks the symmetry and therefore, one has to distinguish between the $"+"$ and $"-"$ directions of the bias magnetic field $B_{bias}$, as indicated in Fig.\ref{fig:Toth_Fig1}(b).

\section{Results and discussions}
\label{Results}

\begin{figure}
\centerline{\includegraphics[width=0.55\textwidth]{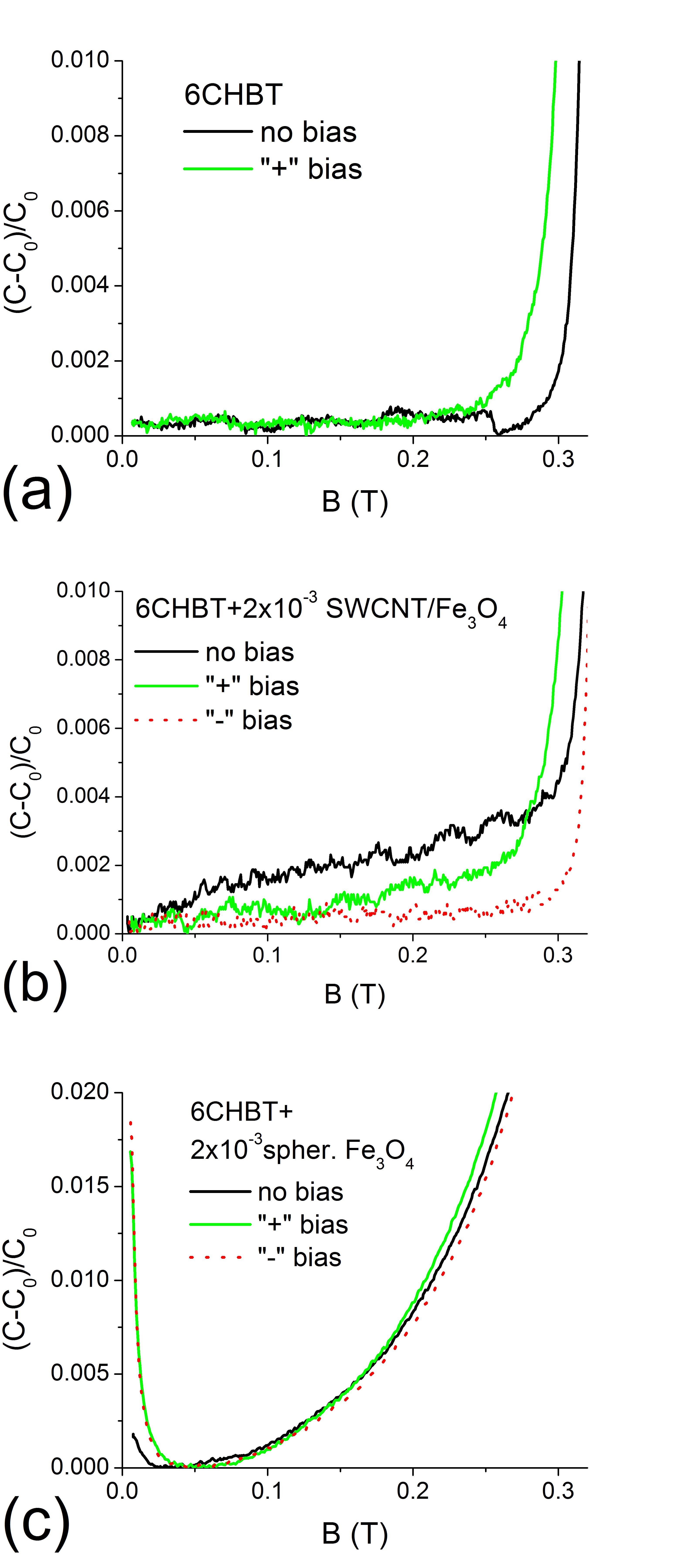}}
\caption{The magnetic field dependence of the relative capacitance measured at $T=30^{\circ}\mathrm{C}$ for 6CHBT (a), 6CHBT doped with SWCNT/Fe$_3$O$_4$ (b), and 6CHBT doped with spherical Fe$_3$O$_4$ nanoparticles (c).}
\label{fig:Toth_Fig2}
\end{figure}

The magnetic field dependence of the relative capacitance variation $(C-C_0)/C_0$ is shown in Fig.\ref{fig:Toth_Fig2} ($C_0$ is the smallest value of the capacitance) with and without a bias magnetic field of $B_{bias}=2$ mT. For undoped 6CHBT neither $B$ nor $B_{bias}$ gave rise to a change of $(C-C_0)/C_0$ below $B_F$ [see Fig.\ref{fig:Toth_Fig2}(a)]. Note that because of the presence of the pretilt, the Fr\'eedericksz transition is not sharp; it becomes continuous in all experiments and therefore, one can define an apparent value of $B_F$ only -- see e.g., Ref.\cite{Meyer1975}. The application of $"+"$ $B_{bias}$ slightly decreases this apparent $B_F$. This is rather surprising, since naively one would expect that $B_{bias}$ stabilizes the initial planar alignment (because of the positive anisotropy of the diamagnetic susceptibility of 6CHBT), and therefore, slightly increases $B_F$. We will come back to this question in a later discussion.

For 6CHBT doped with SWCNT/Fe$_3$O$_4$, a linear dependence of $(C-C_0)/C_0$ on $B$ has been detected below $B_F$ in the absence of $B_{bias}$ [see Fig.\ref{fig:Toth_Fig2}(b)]. The application of $B_{bias}$ of either $"+"$ or $"-"$ directions suppresses this dependence (especially for the $"-"$ direction). This conclusion has also been confirmed by optical measurements of the phase shift $\Delta\varphi$ between the ordinary and extraordinary waves to be discussed later. Note that $"+"$ $B_{bias}$ slightly decreases $B_F$ again (as in 6CHBT), while on the contrary, $"-"$ $B_{bias}$ slightly increases $B_F$ compared to that detected in the absence of $B_{bias}$.

In 6CHBT doped with spherical Fe$_3$O$_4$ nanoparticles the dependence $(\frac {C-C_0} {C_0}) (B)$ is qualitatively different: it is not a monotonic function, but it has a minimum below $B_F$ [see Fig.\ref{fig:Toth_Fig2}(c)]. The Fr\'eedericksz transition becomes "smoother", i.e., the transition is much more continuous than in 6CHBT or in 6CHBT doped with SWCNT/Fe$_3$O$_4$ [{\it cf.} Figs.\ref{fig:Toth_Fig2}(a), (b) and (c)]. On the other hand, $"+"$ and $"-"$ $B_{bias}$ decreases and increases $B_F$, respectively, in a similar manner as in 6CHBT or in the ferronematic with SWCNT/Fe$_3$O$_4$.

The decrease or increase of the apparent $B_F$ depending on the application of $"+"$ or $"-"$ $B_{bias}$, respectively, can be understood by  taking into account the pretilt angle. From the schematic representation in Fig.\ref{fig:Toth_Fig1}(b) it becomes obvious that when both $B$ and $"+"$ $B_{bias}$ are applied, the direction of the net magnetic field encloses a smaller angle with
${\bf n}$ compared to the situation when only $B$ is applied. That leads to a slight decrease of $B_F$ in the former case.
On the contrary, when $B$ and $"-"$ $B_{bias}$ are applied simultaneously the direction of the resulting magnetic field encloses a larger angle with ${\bf n}$ (closer to $90^{\circ}$) leading to an increase of $B_F$.

In the case of the nematic 6CHBT, the effect of the pretilt angle $\theta_0$ can also be discussed more quantitatively if one considers the basic magnetic properties of LCs.
The magnetic moment $\mathbf{M}$ per volume induced in the nematic LC by an external magnetic field $\mathbf{H}$ is
\begin{equation}
\mathbf{M} = \mathbf{\chi} \mathbf{H},
     \label{eq:01}
\end{equation}
where the diamagnetic susceptibility tensor $\mathbf{\chi}$ is constituted from an isotropic part and from an anisotropic contribution defined by
\begin{equation}
\chi_a = \chi_{\parallel} - \chi_{\perp}
     \label{eq:02}
\end{equation}
($\chi_{\parallel}$ and $\chi_{\perp}$ are the magnetic susceptibilities measured by a magnetic field parallel and perpendicular to $\mathbf n$, respectively)
-- see e.g., Ref.\cite{Stannarius2014}. In terms of the net magnetic induction $\mathbf{B}_n$, the expression for the magnetic moment becomes
\begin{equation}
\mathbf{M} = {\frac{\chi_{\perp}} {\mu_0}}\mathbf{B}_n + {\frac{\chi_a} {\mu_0}} (\mathbf{B}_n\cdot\mathbf{n})\mathbf{n},
     \label{eq:03}
\end{equation}
with $\mu_0$ being the vacuum permeability. The torque $\mathbf{\Gamma}$ exerted on $\mathbf{n}$ by the external magnetic field can be calculated from
\begin{equation}
\mathbf{\Gamma} = \mathbf{M} \times \mathbf{B}_n.
     \label{eq:04}
\end{equation}
From the experimental geometry depicted in Fig.\ref{fig:Toth_Fig1} [taking the rubbing direction along {\it x}, and $\mathbf{B}$ parallel with {\it z}], the initial condition for the director $\mathbf{n}$ is
\begin{equation}
n_x = \cos{\theta_0},  \hspace{5pt} n_y = 0, \hspace{5pt} n_z = \sin{\theta_0}.
     \label{eq:05}
\end{equation}
For the net magnetic induction $\mathbf{B}_n$ without the bias magnetic field ($B_{bias} = 0$)
\begin{equation}
B_x = 0,  \hspace{5pt} B_y = 0, \hspace{5pt} B_z = B,
     \label{eq:06}
\end{equation}
while with the $"+"$ $B_{bias}$ bias magnetic field one has
\begin{equation}
B_x = B_{bias},  \hspace{5pt} B_y = 0, \hspace{5pt} B_z = B,
     \label{eq:07}
\end{equation}
and with the $"-"$ $B_{bias}$ bias magnetic field
\begin{equation}
B_x = - B_{bias},  \hspace{5pt} B_y = 0, \hspace{5pt} B_z = B.
     \label{eq:08}
\end{equation}
With these conditions, calculations for the magnetic torques $\mathbf{\Gamma}_0$, $\mathbf{\Gamma}_+$ and $\mathbf{\Gamma}_-$ without $B_{bias}$, with $"+"$ $B_{bias}$ and with $"-"$ $B_{bias}$, respectively, give:
\begin{equation}
\Gamma_{0x} = 0,  \hspace{5pt} \Gamma_{0y} = -\frac{\chi_a B^2\sin{2 \theta_0}}{2 \mu_0}, \hspace{5pt} \Gamma_{0z} = 0,
     \label{eq:09}
\end{equation}
\begin{equation}
\Gamma_{+x} = 0,  \hspace{5pt} \Gamma_{+y} = -\frac{\chi_a[(B^2-B_{bias}^2)\sin{2 \theta_0} + 2 B B_{bias}\cos{2 \theta_0} ]}{2\mu_0}, \hspace{5pt} \Gamma_{+z} = 0,
     \label{eq:010}
\end{equation}
\begin{equation}
\Gamma_{-x} = 0,  \hspace{5pt} \Gamma_{-y} = -\frac{\chi_a[(B^2-B_{bias}^2)\sin{2 \theta_0} - 2 B B_{bias}\cos{2 \theta_0} ]}{2\mu_0}, \hspace{5pt} \Gamma_{-z} = 0.
     \label{eq:011}
\end{equation}
Obviously, for the experimental conditions depicted in Fig.\ref{fig:Toth_Fig1}(b) ($B \gg B_{bias}$ and $\theta_0$ is of a few degrees): $|\Gamma_{+x}| > |\Gamma_{0x}| > |\Gamma_{-x}|$, i.e., the magnetic torque acting on the director is the largest with $"+"$ $B_{bias}$, while with $"-"$ $B_{bias}$ it is the smallest. Similar calculation for the ferronematics is far more complicated, since then the magnetic moments of the magnetic particles as well as the anchoring energy at the surface of the particles has to be taken into account \cite{Burylov1994}.

In parallel with the dielectric studies, the optical phase shift $\Delta\varphi$ between the ordinary and extraordinary waves was  determined from the magnetic field dependence of the light intensity $I$ transmitted through the cell between crossed polarizers using the relation:
\begin{equation}
I = I_0 \sin^2\left(\frac {\Delta\varphi} {2} \right) \sin^2 2\alpha ,
     \label{eq:1}
\end{equation}
where: $I_0$ is the incident light intensity, $\alpha=45^{\circ}$ is the angle between the polarizer and the initial director ${\bf n}$ -- see e.g., \cite{Podoliak2012,Majumdar2011}.

\begin{figure}
\centerline{\includegraphics[width=0.6\textwidth]{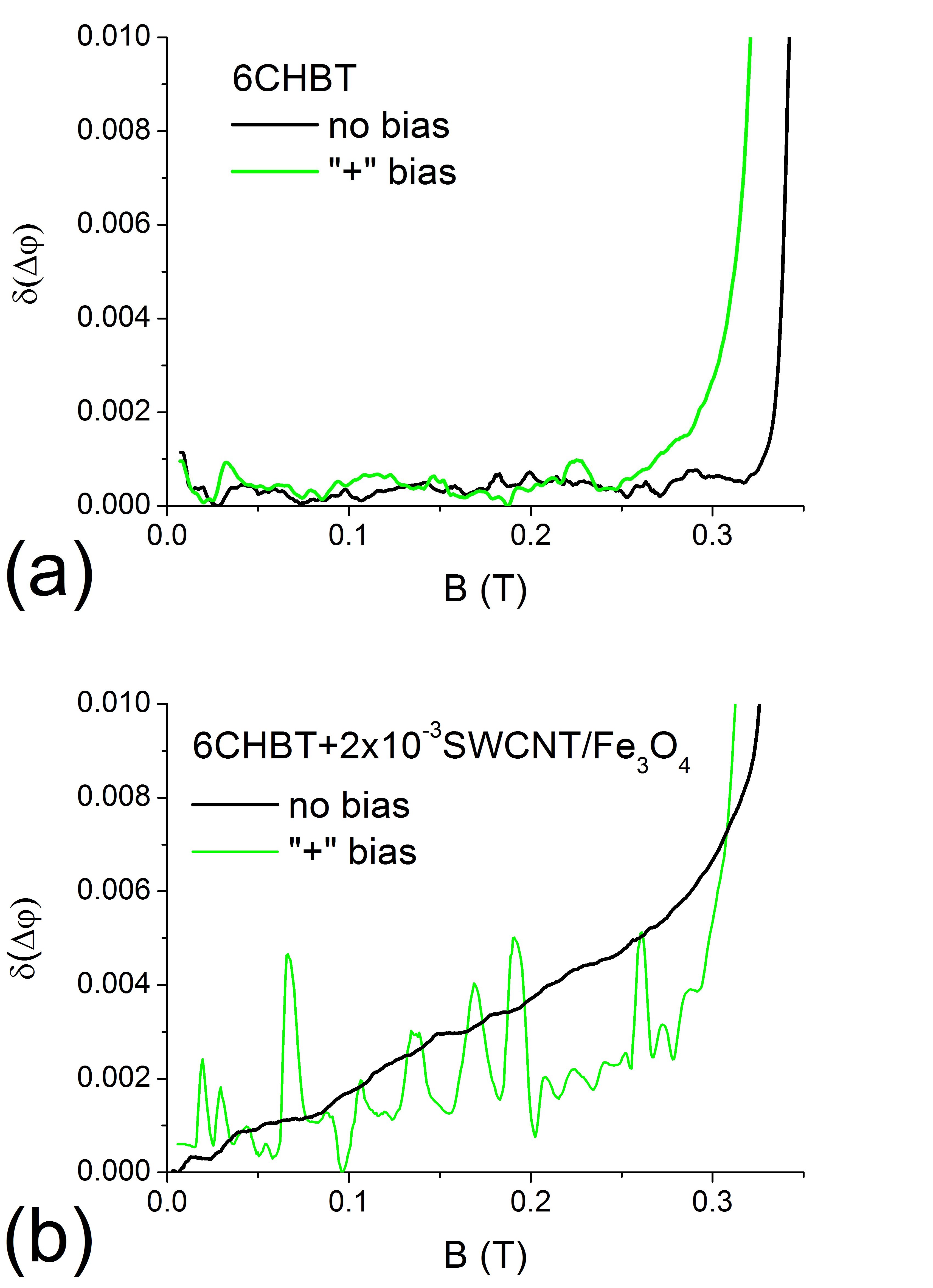}}
\caption{The magnetic field dependence of the relative change in the phase shift $\delta (\Delta \varphi)$ measured at $T=30^{\circ}\mathrm{C}$ in 6CHBT (a), and in 6CHBT doped with SWCNT/Fe$_3$O$_4$ (b) with or without $"+"$ $B_{bias}$ as indicated in the legend.}
\label{fig:Toth_Fig3}
\end{figure}

For an easier comparison with the dielectric data in Figs.\ref{fig:Toth_Fig2}(a) and (b), and in accordance with Figs.2 and 3 of Ref.\cite{Podoliak2011} and with Fig.7 of Ref.\cite{Podoliak2012}, in Fig.\ref{fig:Toth_Fig3} we plot the relative change in the phase shift $\delta(\Delta\varphi)$ defined as
\begin{equation}
\delta(\Delta\varphi) = \frac{\Delta\varphi_0 - \Delta\varphi}{\Delta\varphi_0}
     \label{eq:2}
\end{equation}
(where $\Delta\varphi_0$ and $\Delta\varphi$ are the phase shifts for $B=0$ and $B\neq0$, respectively) as a function of the magnetic induction $B$ for both 6CHBT and 6CHBT doped with SWCNT/Fe$_3$O$_4$, with and without the $"+"$ $B_{bias}$.

The optical measurements presented in Fig.\ref{fig:Toth_Fig3} support the results of the dielectric studies. For 6CHBT the phase shift does not depend on $B$ below $B_F$ and the $"+"$ $B_{bias}$ decreases the value of $B_F$ -- see Fig.\ref{fig:Toth_Fig3}(a).
For 6CHBT doped with SWCNT/Fe$_3$O$_4$ the dependence of $\delta(\Delta\varphi)$ on $B$ is linear below $B_F$ [Fig.\ref{fig:Toth_Fig3}(b)]. When $"+"$ $B_{bias}$ was applied, though the response became more noisy, evidently it is much smaller than without a bias magnetic field, i.e., $B_{bias}$ suppresses the low magnetic field effect similarly to what is obtained by the capacitance measurements [Fig.\ref{fig:Toth_Fig2}(b)]. Again, the apparent value of $B_F$ is slightly decreased when the $"+"$ $B_{bias}$ is applied [Fig.\ref{fig:Toth_Fig3}(b)].

\begin{figure}
\centerline{\includegraphics[width=0.6\textwidth]{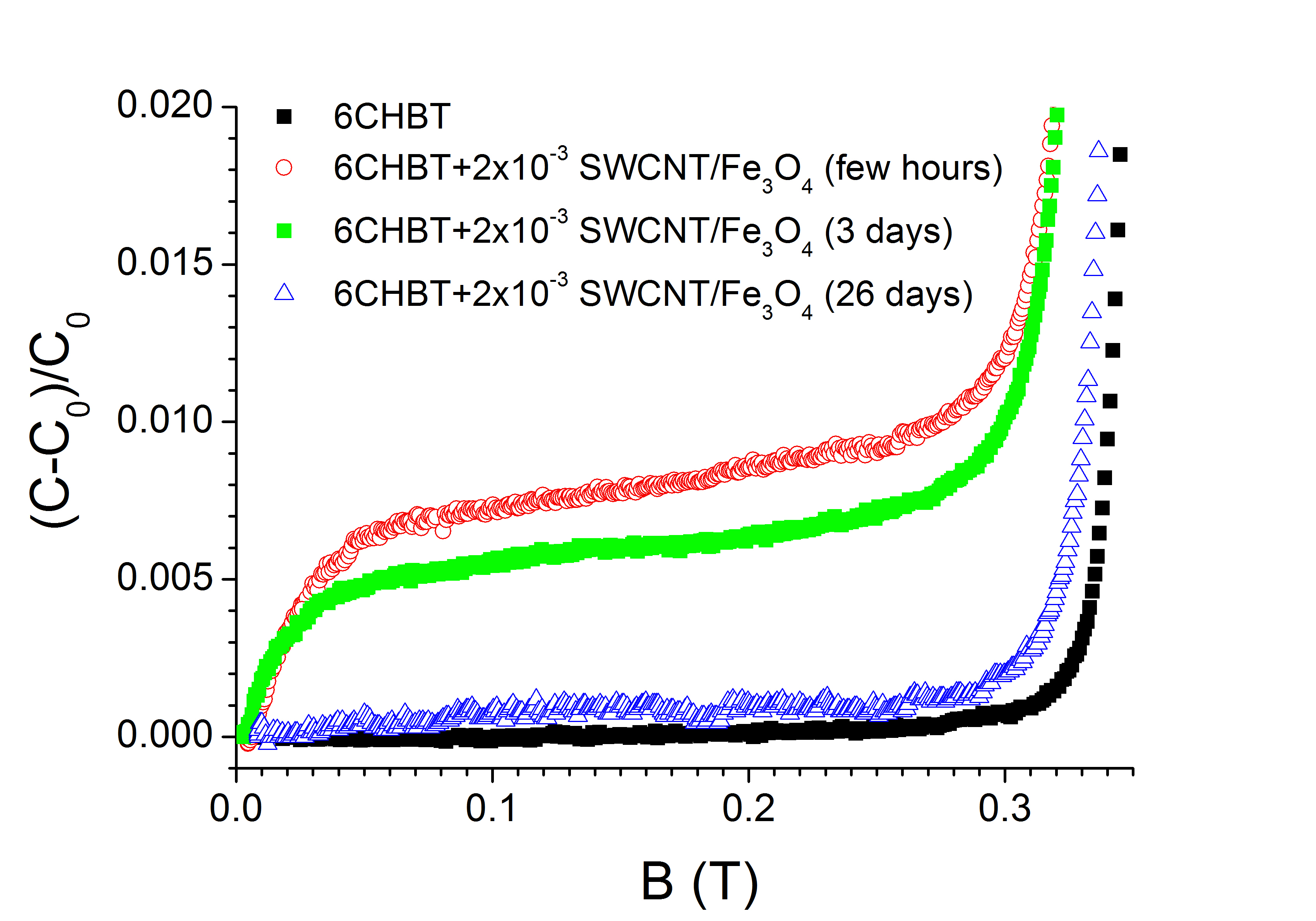}}
\caption{The magnetic field dependence of the relative capacitance measured at $T=25^{\circ}\mathrm{C}$ for 6CHBT and 6CHBT doped with SWCNT/Fe$_3$O$_4$ measured at different times elapsed from the cell preparation.}
\label{fig:Toth_Fig4}
\end{figure}

Another focus of the present work was to investigate how the aggregation of nanoparticles influences the response of the ferronematics to low magnetic fields. For this purpose, a sample of 6CHBT and a cell filled with 6CHBT doped with SWCNT/Fe$_3$O$_4$ has been monitored on a long time scale without a bias magnetic field $B_{bias}$.
We measured the relative capacitance versus B after different times elapsed from the cell preparation in a ferronematic system with carbon nanotubes and compared them with the time independent characteristics of 6CHBT. The results are presented in Fig.\ref{fig:Toth_Fig4}.

As one sees, the first measurement on the ferronematic (made a few hours after its preparation) results in the largest capacitive response to the applied magnetic field $B$. As time elapsed, the response got weaker, and within a month it almost disappeared: after 26 days from preparation the response of the ferronematic differs from that of 6CHBT only in small details (a slight slope of $(\frac {C-C_0} {C_0}) (B)$, and a somewhat smaller $B_F$).

\begin{figure}
\centerline{\includegraphics[width=0.6\textwidth]{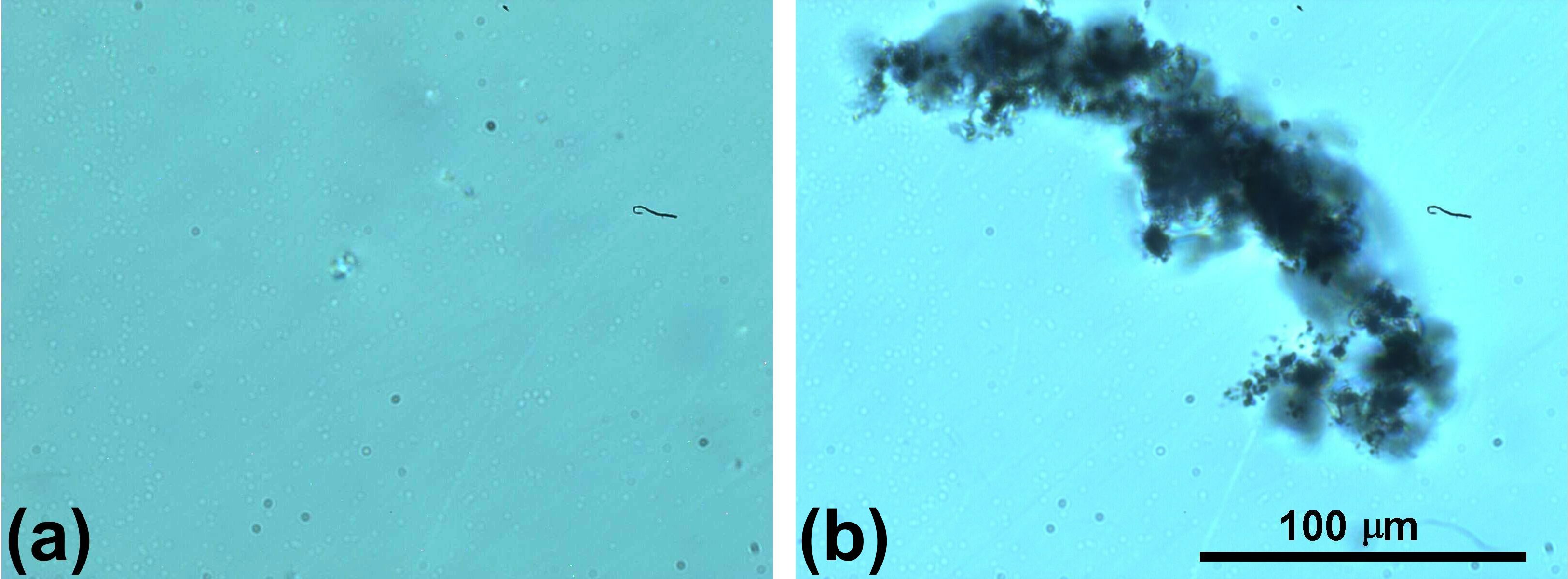}}
\caption{Microscopic images taken about four months after the sample preparation on a cell filled with 6CHBT (a), and on a cell with 6CHBT doped with SWCNT/Fe$_3$O$_4$ (b). The magnification of the subfigures is the same.}
\label{fig:Toth_Fig5}
\end{figure}

The idea that the aggregation of nanoparticles is behind the above described effect is supported by optical microscopy. Fig.\ref{fig:Toth_Fig5} shows pictures taken by a polarizing microscope on 6CHBT (a), and on 6CHBT doped with SWCNT/Fe$_3$O$_4$ (b) four months after the sample preparation. Obviously, in the ferronematic nanoparticle aggregates of the size of the order of tens of micrometer are observable.

\begin{figure}[h]
\centerline{\includegraphics[width=0.6\textwidth]{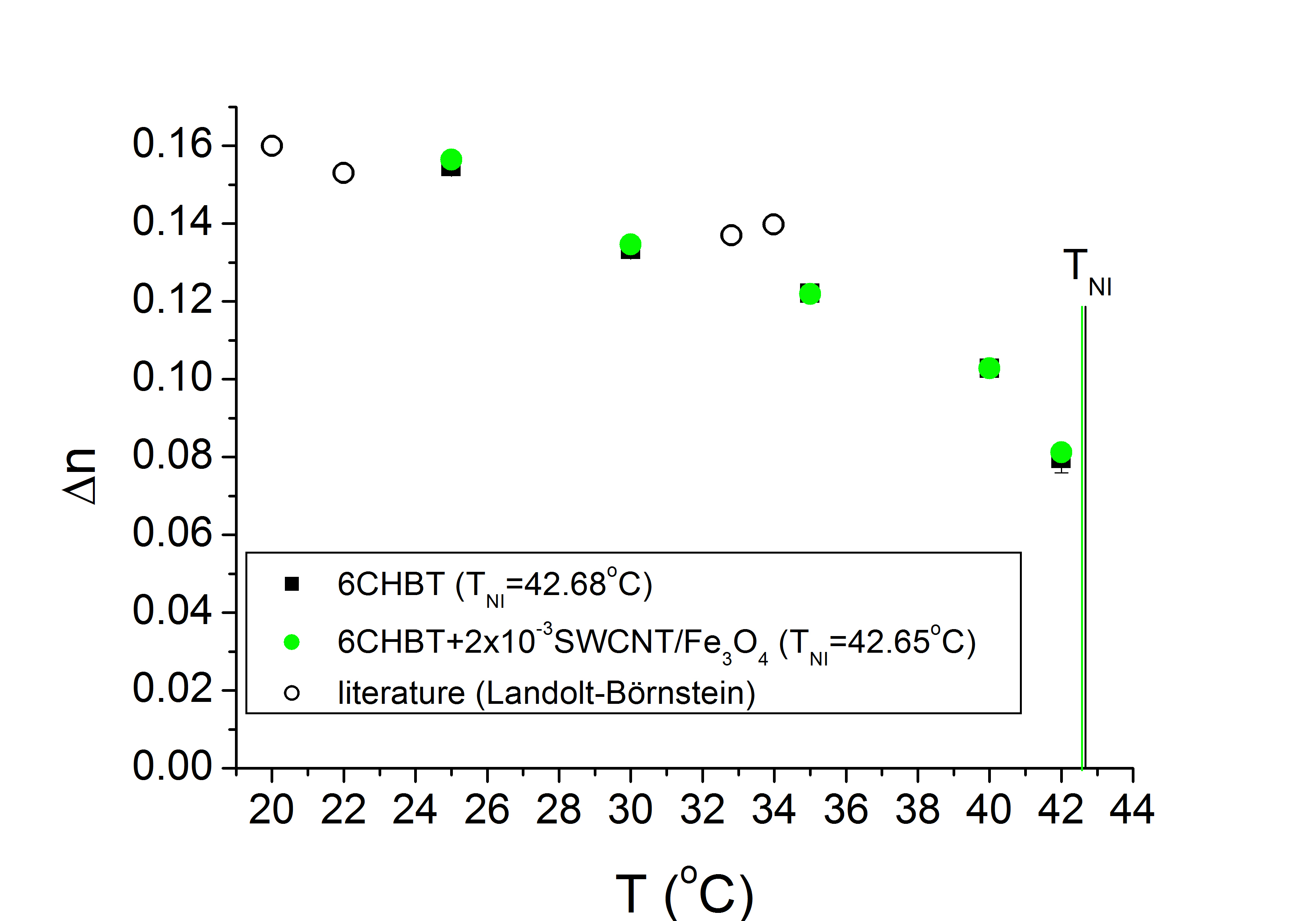}}
\caption{Temperature dependence of the birefringence $\Delta n$ measured for 6CHBT and 6CHBT doped with SWCNT/Fe$_3$O$_4$ compared to the values for 6CHBT taken from the literature \cite{Landolt}.}
\label{fig:Toth_Fig6}
\end{figure}

Lastly we present results on the temperature dependence of the birefringence $\Delta n$.
Using the measured maximal phase shift $\Delta\varphi_0$ and sample thickness $d$ for a known $\lambda$, the birefringence $\Delta n$ can be calculated. In Fig.\ref{fig:Toth_Fig6} the temperature dependence of the birefringence is presented for 6CHBT as well for 6CHBT doped with SWCNT/Fe$_3$O$_4$. Data taken from the literature \cite{Landolt} are also shown for comparison. From Fig.\ref{fig:Toth_Fig6} several conclusions can be made. First, doping 6CHBT with SWCNT/Fe$_3$O$_4$ even in a relatively high concentration does not influence significantly the nematic to isotropic phase transition temperature $T_{NI}$. Secondly, the doping does not change the birefringence, and thirdly, our results are in reasonable agreement with the data from the literature. Finally, we mention that in order to obtain precise values of $\Delta n$, one has to achieve a full realignment during the Fr\'eedericksz transition; i.e., one has to increase the field to several times of the threshold value. Due to limitations of our electromagnet a high enough magnetic field could not be reached ($B_{max}\approx 1$~T corresponds to $\approx 3.3 B_F$). Therefore, the electric field induced Fr\'eedericksz transition was used for the $\Delta n$ measurements.

\section{Conclusion}
\label{Conclusion}

In summary, we have shown that the orienting bias magnetic field $B_{bias}$ is not a prerequisite for the response of ferronematics to low magnetic fields. Moreover, as we have demonstrated, in some cases $B_{bias}$ even suppresses the response.
On the other hand, $B_{bias}$ shifts the critical field of the magnetic Fr\'eedericksz transition $B_F$ (increases or decreases it depending on the direction of $B_{bias}$) because of the presence of a pretilt in planarly oriented samples. We have pointed out the importance of the aggregation of nanoparticles, which decreases the response of ferronematics to low magnetic fields. We have also shown that doping the LC with SWCNT/Fe$_3$O$_4$ does not change the birefringence, nor the nematic to isotropic phase transition temperature.
The experimental results presented in this work give rise to further questions for which the answers require additional experimental and theoretical research in the future. Among these questions we underline two. First, it is still unknown what and by which mechanism(s) causes the optical and dielectric response of ferronematics to low magnetic fields. Second, why is this response qualitatively so much different in ferronematics obtained by doping with SWCNT/Fe$_3$O$_4$ and with spherical Fe$_3$O$_4$
[{\it cf.} Figs.\ref{fig:Toth_Fig2}(b) and (c)]? The latter question is even more intriguing in the light of the results obtained for lower ($\leq10^{-3}$) volume concentrations of the spherical Fe$_3$O$_4$ nanoparticles and at somewhat higher temperature ($T=35^{\circ}\mathrm{C}$), where a linear $(\frac {C-C_0} {C_0}) (B)$ has been obtained \cite{Tomasovicova2013} in contrast to the non-monotonic behavior shown in Fig.\ref{fig:Toth_Fig2}(c).

\section*{Acknowledgements}
\label{Acknowledgements}

Financial support by the Hungarian Research Fund OTKA K81250, by FP7 M-Era.Net 2012 MACOSYS (OTKA NN110672), and by the Ministry of Education Agency for Structural Funds of EU in the frame of the project 26220120021 is gratefully acknowledged.
T.T.-K. and N.\'E. are thankful for the hospitality provided in the framework of the HAS-SAS Bilateral Mobility Grant.

\end{document}